\documentclass[a4paper,10pt]{article}

\usepackage{times,fullpage}
\usepackage{mathtools,amsfonts,amssymb,amsmath} 
\usepackage{slashed}
\usepackage[affil-it]{authblk} % enables the address as \affil in the title
\usepackage{mygraphs} % this is my package of rules of Feynman graphs drawings

\begin{document}

\renewcommand{\thefootnote}{\fnsymbol{footnote}}

\begin{titlepage} % TITLE PAGE
\title{Doublet-singlet model and unitarity}
\author{ G. Cynolter\footnote{cyn@general.elte.hu}, \, 
J. Kov\'{a}cs\footnote{kovacsjucus@caesar.elte.hu} \, and 
E. Lendvai\footnote{lendvai@general.elte.hu} }
\affil{MTA-ELTE Research Group in Theoretical Physics, E\"otv\"os University,
Budapest, 1117 P\'azm\'any P\'eter s\'et\'any 1/A, Hungary}
\date{} % the empty date takes out the date

\maketitle
\thispagestyle{empty}

\begin{abstract}
We study the renormalizable singlet-doublet fermionic extension of the Standard Model. In this model, the new vector-like fermions couple to the gauge bosons and to the Higgs via new Yukawa couplings, that allow for nontrivial mixing in the new sector, providing  a stable, neutral dark matter candidate. Approximate analytic formulae are given for the mass spectrum around the blind spots, where the dark matter candidate coupling to $h$ or $Z$ vanishes. We calculate the two particle scattering amplitudes in the model, impose the perturbative unitarity constraints and establish bounds on the Yukawa couplings.
\end{abstract}

%\tableofcontents
\end{titlepage}

\renewcommand{\thefootnote}{\arabic{footnote}}

\section{Introduction} % INTRO

There is a renewed interest in vector-like fermions \cite{survey}, as these are less constrained than chiral fermions in electroweak precision tests of the Standard Model (SM) \cite{cynprec,schwaller} and free of anomalies. Vector-like fermions appear naturally in various beyond the Standard Model scenarios, such as specific string theories, bulk femions in universal extra dimensional models \cite{UED}, little Higgs theories \cite{lH,lH2}, supersymmetric models as superpartners of standard bosons e.g. higgsinos \cite{lowSUSY}, various composite Higgs models \cite{Contino,fcm} and simplified models of dark matter \cite{DMandUnif,Hall,DEramo}. Vector-like fermions help gauge coupling unification, but the lower unification scale is in conflict with proton decay constraints \cite{DMandUnif}. A recent survey of phenomenological implications of vector-like extensions \cite{survey} presents constraints from electroweak precision observables, direct collider searches, Higgs production and decays. 

The couplings to the standard light (and heavier) fermions are severely constrained. To avoid flavour problems (for flavour issues see \cite{ligeti}) and to ensure there is no mixing between SM and the new vector-like fermions a $\mathbb{Z}_2$ matter parity is introduced. The new fermions are odd, while the standard particles are even under this symmetry. The lightest new particle is stable and if it is electrically neutral then it provides a dark matter candidate. The singlet-doublet dark matter model is an effective theory or simplified version of the neutralino MSSM dark matter sector. The mixing in the dark sector allows for a wide range of Higgs and gauge boson couplings that is able to avoid direct detection in dark matter experiments and colliders and give the measured relic density, while still remaining a relatively simple model. The model contains four new parameters, two dimensionful mass parameters and two Yukawa couplings, originally completely unconstrained. A recent analysis in \cite{SDbounds} has found combined bounds on the vector masses and the overall value of the Yukawa coupling.

The spectrum generally is a solution of a third order equation and the  analytical treatment is difficult. Another problem is that the four-dimensional parameter space can be only visualized if at least two parameters (generally the Yukawas) are fixed. In  \cite{SDbounds} the phenomenologically allowed regions of the parameter space have been found in the neighborhood of the blind spots, where the coupling of the dark matter candidate to $h$ or $Z$ vanishes. In these special points one eigenvalue can be found and the third order equation simplifies to a second order one. With this observation we have found analytic solution for the masses close to the blind spots. We have further shown that the relevant coupling constants are small and can avoid observation by direct detection experiments.

The parameters of the model can be constrained on theoretical grounds. Perturbative unitarity is a useful tool to set limits on effective theories. Starting from weak charged current interaction, it leads to the well-known gauge bosons and the Higgs with usual couplings of the SM. It provided an upper bound on its validity without the Higgs boson and a theoretical upper bound on the Higgs mass \cite{BenLee}. For chiral fermions, there is also an upper bound on the scale of fermion mass generation in the few TeV region \cite{appel}. In this paper we calculate the two-particle scattering amplitudes involving the new fermions, aiming to constrain the free parameters of the model. As the new part of the Lagrangian is renormalizable, the potentially dangerous amplitudes growing with energy cancel each other. The elastic scattering of neutral fermions of the doublet and the singlet gives a meaningful amplitude, constraining the value of new Yukawa couplings. The new vector-like fermions only partially receive their masses from the Higgs, therefore the new bound, 
$ \frac{|y_{1,2}|v}{\sqrt{2}} \leq 1.23  $ TeV
cannot be directly translated to the mass of the dark matter candidate or to the mass parameters of the model.

\par In section \ref{sec:models} we review the singlet-doublet vector-like fermion extension of the Standard Model, then comment on the current dark matter bounds. In section \ref{sec:expansion} we give analytic formulae around the experimentally favored regions. In section \ref{sec:unitarity} we calculate the new two-particle scattering processes that can contribute to bound the model parameters from perturbative unitarity. The paper is closed with conclusion.

\section{The model} \label{sec:models} % MODELS

We extend the Standard Model with a pair of $SU(2)_W$ doublet Weyl-fermions, $\psi_1=\begin{pmatrix}\psi_1^0 \\ \psi_1^-\end{pmatrix}$ and $\psi_2=\begin{pmatrix}\psi_2^+ \\ \psi_2^0\end{pmatrix}$, that acquire a Dirac mass term together and a singlet, $\chi^0$ with a Majorana mass term. All three are color-singlet to avoid the strong collider bounds, their quantum numbers are listed in Table \ref{tab:QNs}. We also assume a matter parity-like $\mathbb{Z}_2$ symmetry that forbids the new fermions to couple directly to the SM ones. The new particles then only couple to the Higgs and gauge bosons in pairs, so the Lagrangian can be break up into two parts, $\mathcal{L}=\mathcal{L}_{SM}+\mathcal{L}_{DS}$.

\begin{table} [ht] % t(top)b(bottom)p(float page) is the default, h is for here
\centering 
\begin{tabular}{|c|c|c|c|}
\hline
 & $T_3$ & $Q$ & $Y=Q-T_3$ \\ \hline
$\psi_1^0$ & $\phantom{-}\frac{1}{2}$ & $\phantom{-}0$ & $-\frac{1}{2}$ \\ \hline
$\psi_1^-$ & $-\frac{1}{2}$ & $-1$ & $-\frac{1}{2}$ \\ \hline
$\psi_2^+$ & $\phantom{-}\frac{1}{2}$ & $+1$ & $\phantom{-}\frac{1}{2}$ \\ \hline
$\psi_2^0$ & $-\frac{1}{2}$ & $\phantom{-}0$ & $\phantom{-}\frac{1}{2}$ \\ \hline
$\chi^0$ & $\phantom{-}0$ & $\phantom{-}0$ & $\phantom{-}0$ \\ \hline
$h$ & $-\frac{1}{2}$ & $\phantom{-}0$ & $\phantom{-}\frac{1}{2}$ \\
\hline
\end{tabular}
\caption{The weak quantum numbers of the new fermions and the Higgs.} \label{tab:QNs}
\end{table}

The Lagrangian for the new fermions contains their kinetic and mass terms and Yukawa couplings. 
\begin{multline}
 \mathcal{L}_{DS}=\frac{i}{2}\left(\chi^{0\dagger}\bar{\sigma}^{\mu}\partial_{\mu}\chi^0+\psi_1^{\dagger}\bar{\sigma}^{\mu}D_{\mu}\psi_1+\psi_2^{\dagger}\bar{\sigma}^{\mu}D_{\mu}\psi_2\right) \\ 
-\left(m_d\psi_1\psi_2+\frac{1}{2}m_s\chi^0\chi^0+y_1\psi_1H\chi^0+y_2\psi_2\tilde{H}\chi^0+h.c.\right)
\end{multline}
There are two new dimensionful mass parameters and two dimensionless Yukawa couplings, $\left(m_d,\, m_s,\, y_1,\, y_2\right)$. 
The three phases of $\psi_{1,2}$ and $\chi^0$ can be fixed by setting three parameters (e.g. $m_d,\, m_s,\, y_1 $) to be real and positive.  We will consider the case where $y_{2}$ is real\footnote{It can have a CP-violating phase, which is discussed in \cite{DMandUnif}.}
and emphasize that the sign of $y_2/y_1$ is physical.

A nice and widely used parametrization of the Yukawas is  
\begin{equation}
\begin{matrix} y_1=y\cos{\theta} \, ,\, & y_2=y\sin{\theta}. \end{matrix}
\end{equation}
The decoupled MSSM bino-higgsino system with one light Higgs corresponds to 
$\beta =\theta$, and $y$ is related to the $U(1)$ gauge coupling.
\par The charged fermions, $\psi_1^-$ and $\psi_2^+$ merge into a Dirac fermion, $\Psi^-=\begin{pmatrix}\psi_1^- \\ \psi_2^{+\dagger}\end{pmatrix}$ with mass $m_d$. Without the Yukawa couplings, the neutral part of the doublets also form a Dirac fermion with mass $m_d$, $\Psi^0=\begin{pmatrix}\psi_1^0 \\ \psi_2^{0\dagger}\end{pmatrix}$. As the Higgs gets a vacuum expectation value with non-vanishing Yukawa couplings, there is a mixing in the new neutral sector. \begin{equation}
\mathcal{L}_{DS}\supset-\frac{1}{2}
\begin{pmatrix}\chi^0 & \psi_1^0 & \psi_2^0 \end{pmatrix}M_n
\begin{pmatrix}\chi^0 \\ \psi_1^0 \\ \psi_2^0 \end{pmatrix}+h.c.,
\end{equation}
with the mass matrix $M_n$,
\begin{equation}
M_n=\begin{pmatrix}
m_s & \frac{y_1 v}{\sqrt{2}} & \frac{y_2 v}{\sqrt{2}} \\
\frac{y_1 v}{\sqrt{2}} & 0 & m_d \\ \frac{y_2 v}{\sqrt{2}} & m_d & 0 
\end{pmatrix}.
\end{equation}
The corresponding characteristic equation follows, 
\begin{equation}
\left(m_s-\lambda\right)\left(\lambda^2-m_d^2\right)+m_dy_1y_2v^2+\lambda\frac{\left(y_1^2+y_2^2\right)v^2}{2}=0.
\label{eq:char}
\end{equation}
\par The cubic equation can be solved analytically using the Cardano formula or numerically. There is a negative eigenvalue that can be flipped to be positive, multiplying the corresponding eigenvector by $i$ or equivalently performing the Takagi diagonalization on the system and get only the positive masses. Generally the spectrum contains three neutral Majorana fermions, the mass eigenstates $\chi_1,\,\chi_2,\,\chi_3$. The lightest will be denoted by $\chi$ with mass $m_{\chi}$, that will be stable due to the $\mathbb{Z}_2$ symmetry. It is an ideal dark matter candidate if lighter than the charged fermion $m_{\chi}<m_d$, made from the following composition of the weak eigenstates, 
\begin{equation}
\chi=U_{11}\chi^0+U_{12}\psi_1^0+U_{13}\psi_2^0\text{, }\, \, |U_{11}|^2+|U_{12}|^2+|U_{13}|^2=1.
\end{equation}
$U_{11}^2$ characterizes the amount of the singlet in $\chi$ . When $m_d>m_s$, $\chi$ is more singlet- or bino-like with $U_{11}^2>0.5$, while when $m_d<m_s$, $\chi$ is more doublet- or higgsino-like with $U_{11}^2<0.5$.

\section{Dark matter constraints} \label{sec:DM} % DM BOUNDS

The singlet-doublet model for various values of the parameters can provide a wino- or a higgsino-like dark matter candidate.
The parameters of the model and the dark matter candidate were recently explored in \cite{singletdoublet,Higgsoasis,simplModels,Iceberg}.
\par Several experiments constrain dark matter candidates. Planck observations \cite{Planck} on the relic density abundance $\Omega_{dm}h^2\simeq 0.12$. Indirect searches obtained by Fermi-LAT \cite{FermiLAT} and IceCube \cite{IceCube}, and direct dark matter searches by PICO \cite{PICO}, LUX \cite{LUX} and XENON100 \cite{X100}. In the low mass region, collider bounds are becoming important, too, as the charged fermion mass is already excluded for $m_d<100\, \text{GeV}$ by chargino searches at LEP \cite{LEP} and the LHC can study the production and annihilation of the dark matter particle \cite{LHC} or the invisible Higgs and $Z$ decays.
\par The most stringent bounds are coming from the direct detection experiments in the low mass region. Here the relevant couplings of the dark matter are the ones to the Higgs and $Z$. 
It is important to note that the $c_{h\chi\chi}$ coupling is  related to the the characteristic equation. After differentiating \eqref{eq:char} with respect to the Higgs VEV, it can be solved for $\frac{\partial m_{\chi}}{\partial v}=c_{h\chi\chi}$.
\begin{equation}
c_{h\chi\chi}=-\frac{\left(2y_1y_2m_d+\left(y_1^2+y_2^2\right)m_{\chi}\right)v}{m_d^2+\left(y_1^2+y_2^2\right)\frac{v^2}{2}+2m_sm_{\chi}-3m_{\chi}^2},
\label{ch}
\end{equation}
\begin{equation}
c_{Z\chi\chi}=-\frac{m_zv\left(y_1^2-y_2^2\right)\left(m_{\chi}^2-m_d^2\right)}{2\left(m_{\chi}^2-m_d^2\right)^2+v^2\left(4y_1y_2m_{\chi}m_d+\left(y_1^2+y_2^2\right)\left(m_{\chi}^2+m_d^2\right)\right)},
\label{cZ}
\end{equation}
where $m_\chi$ is the eigenvalue, which can be negative as well.

The spin-independent "blind spots" \cite{blindspot} are defined, where $c_{h\chi\chi}$ coupling vanishes, similarly the spin-dependent blind spots, where $c_{Z\chi\chi}$ coupling vanishes. These points evade the related direct detection bounds. To keep the couplings small and avoid overabundance in the relic density, the annihilation cross section needs enhancement. This can be achieved, if there is a pole in the s-channel processes, e.g. $m_{\chi}\approx\frac{m_h}{2}\text{ or }\frac{m_z}{2}$ \cite{singletdoublet}. A recent work \cite{SDbounds} collected all these bounds on the singlet-doublet model, including dark matter searches and colliders. It carries out a full numerical study to cover the parameter space, that includes a higher mass region, where $m_{\chi}>100 \, \text{GeV}$ and a light mass region, both illustrated with three representative Yukawa couplings $y$. 
\par In the large mass region, indirect detection constrains the underabundant region in the parameter space. For the smaller Yukawas $y\leq 0.01$ that is the only bound and $m_{\chi}\gtrsim 280\, \text{GeV}$ with $U_{11}^2\gtrsim 0.5$ remains available for any $\tan{\theta}$. For the more 'MSSM-like' $y=0.2$, it allows $m_{\chi}\gtrsim 220 \, \text{GeV}$ with $U_{11}^2\gtrsim 0.65$. Here the direct detection experiments are relevant for thermal $\chi$ with $\Omega_{\chi}h^2\approx 0.12$, in the region where $m_s\approx m_d$ LUX excludes up to $m_{\chi}\sim 1\, \text{TeV}$ mass. In the third region with larger Yukawas $y=1$, they obtained $m_{\chi}\gtrsim 275 \, \text{GeV}$, unless it is purely singlet ($U_{11}^2\gtrsim 0.8$).

\par In the light mass region  $\chi$ is singlet-like as we mentioned, the allowed regions are around the blind spots  with $m_{\chi}\approx\frac{m_h}{2}\text{ or }\frac{m_z}{2}$, but these are further constrained by the invisible Higgs decay. Parametrically, for the smaller Yukawa coupling $y=0.1$, $\chi$ is excluded in the $80\, \text{GeV}\lesssim m_{\chi}\lesssim 220\, \text{GeV}$ range with $U_{11}^2\lesssim 0.65$, and there is no further bounds on the $y=1$ region.
\par The experiments bound the parameter space, but four parameters are challenging to interpret. In the next section we explore analytically the regions around the blind spots.

\section{Regions of small couplings} \label{sec:expansion} % EXPANSION 

The bounds  from direct detection experiments can be fulfilled with small dark matter couplings to the Higgs and $Z$ bosons (\ref{ch}, \ref{cZ}). The LUX \cite{LUX} and XENON100 \cite{X100} experiments set bounds on the spin independent (and dependent) cross section giving for a dark matter heavier than the nucleon \cite{blindspot} $c_{h\chi \chi} \leq0.01-0.1$ and similarly $c_{Z\chi \chi} \leq0.01-0.1$ depending on the dark matter mass. In this section we study analytically the region around the blind spots, the experimentally favored regions of the parameter space \cite{SDbounds}.

The blind spots are the following,
\begin{equation}
c_{Z\chi\chi}=0, \; \mathrm{if}\; |y_1|=|y_2| \; \mathrm{or} \; |m_\chi|=m_d
\label{eq:Z cond}
\end{equation}
and
\begin{equation}
c_{h\chi\chi}=0, \; \mathrm{if}\;  m_\chi+\sin(2\theta) m_d=0,  
\label{eq:h cond}
\end{equation}
where $\sin(2\theta)=\frac{2y_1 y_2}{y_1^2+y_2^2}$. The conditions in \eqref{eq:Z cond}
are not independent. If $\lambda=\pm m_d$ is an eigenvalue of the characteristic equation \eqref{eq:char}
it follows that  $y_1=\mp y_2 $ or $\tan \theta =\mp 1 $. 
Therefore it is enough to study the cases in the neighborhood of  $y_1=\pm y_2$ and \eqref{eq:h cond}.
Similarly if $c_{h\chi \chi}$ vanishes, it follows from \eqref{eq:char} that the corresponding eigenvalue $m_\chi$ is $\pm m_d$ or $m_s$.
In both cases the cubic equation remarkably simplifies by finding one root ($m_d$ or $m_s$) and the rest is a quadratic equation.
The expansion is performed around these special points.

\subsection*{Small $c_{Z \chi \chi}$, $y_1\simeq-y_2,$ $\tan{\theta} \simeq -1$}
\begin{figure}[h]
\begin{centering}
\includegraphics[height=3cm]{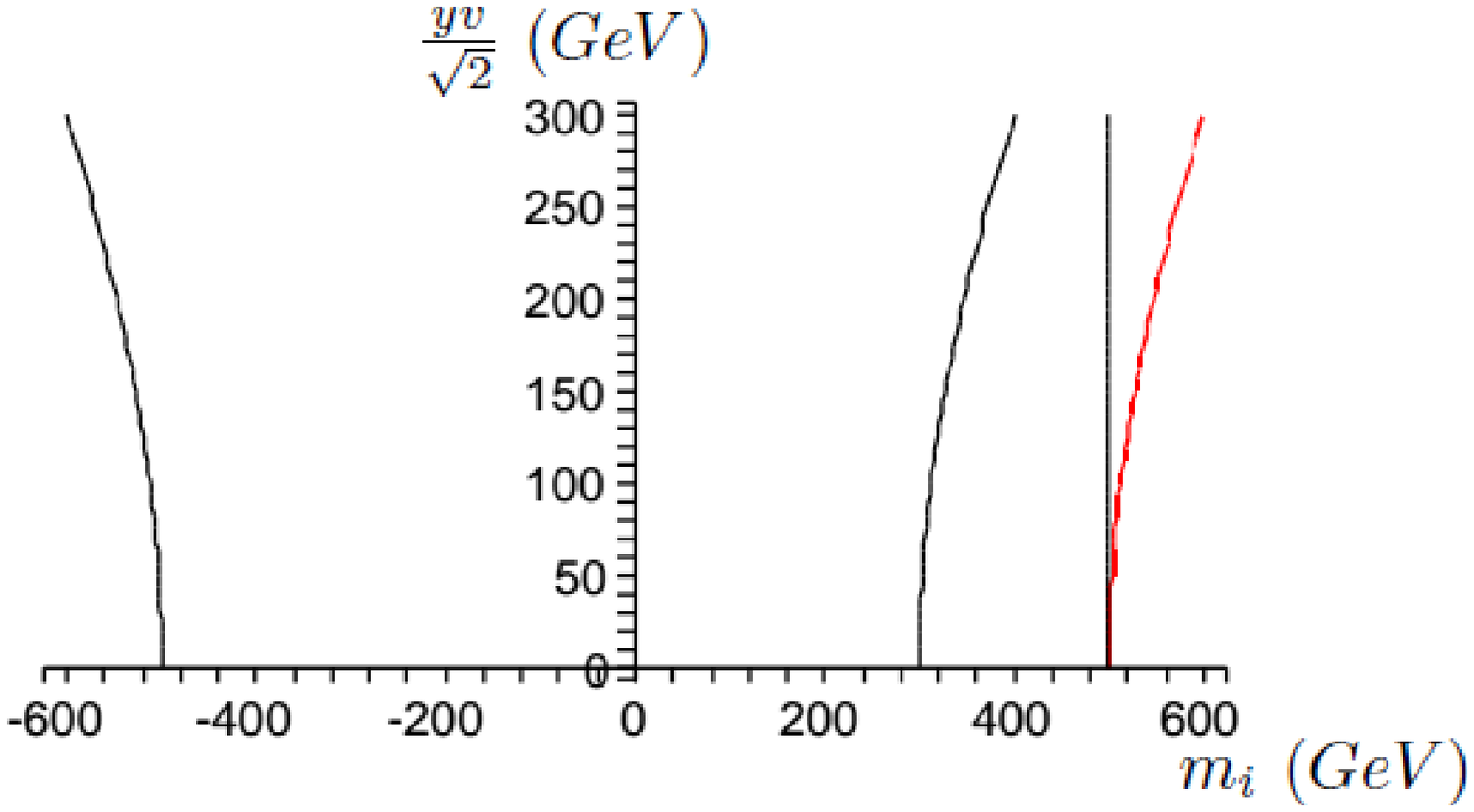}
\includegraphics[height=3cm]{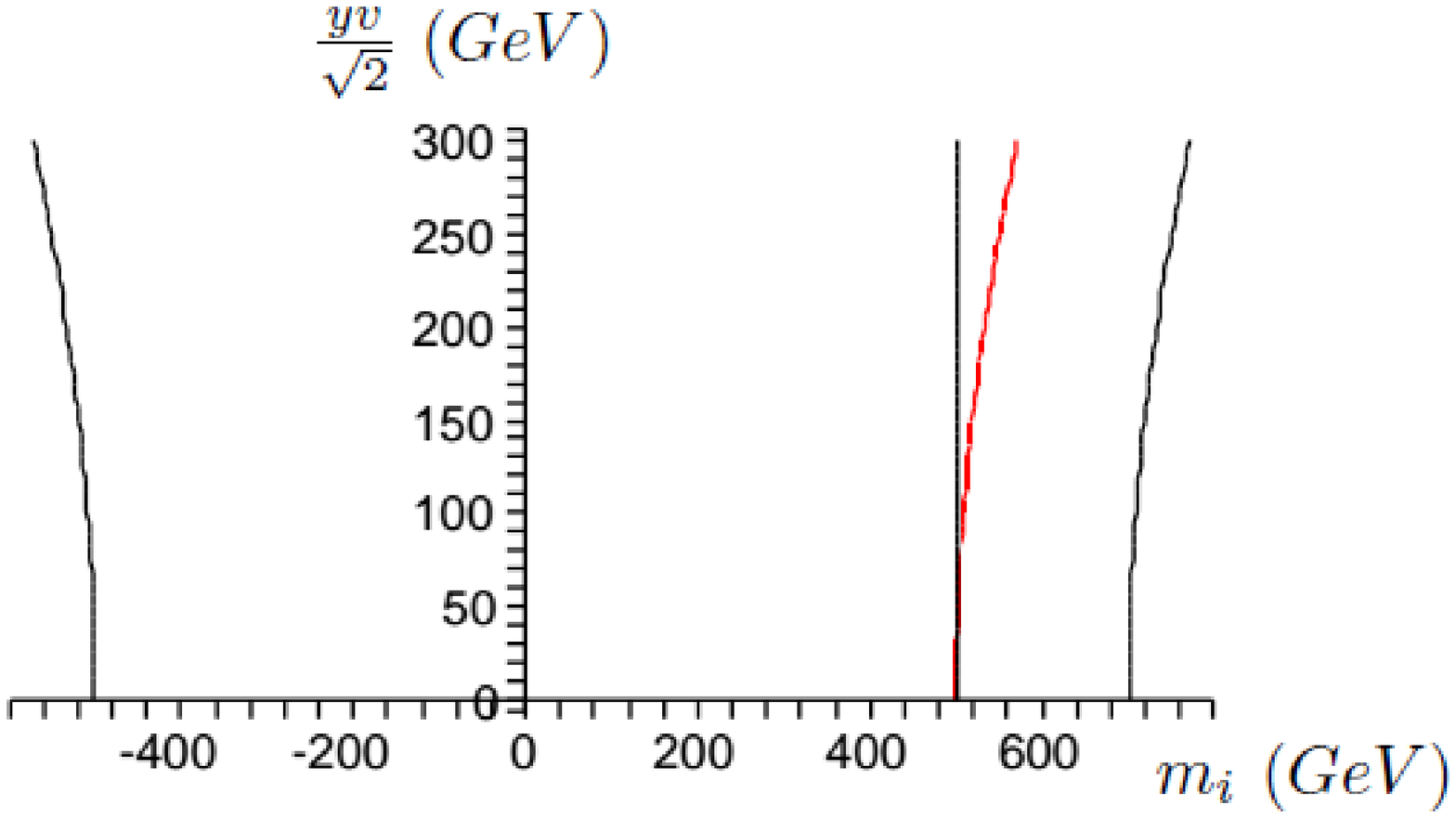}
\par\end{centering}
\centering{}\caption{Mass eigenvalues for $y_1=-y_2$ vs. increasing overall Yukawa terms $\frac{yv}{\sqrt{2}}$ for $m_s=300$ GeV  $<m_d=500$ GeV on the left  and $m_s=800$ GeV  $>m_d=500$ GeV on the right. The negative eigenvalue ($m_3$) has been flipped to positive value and both are shown.}
\label{fig:case1spec}
\end{figure}

The mass eigenvalues are
\begin{equation}
m_1=m_d (1-x_-), \, \quad \mathrm{where} \quad \, x_-=\frac{ (y_1+y_2)^2 v^2 }{ 4 m_d(m_s-m_d)+y^2{v^2} },
\end{equation}
\begin{equation}
m_{2,3}=\frac{m_s-m_d}{2} \pm \frac{1}{2}  \sqrt{(m_s\!+\!m_d)^2+2y^2 v^2}  +
\frac {m_d}{2} x_- \left(  1 \mp \frac{m_s-3m_d}{\sqrt{(m_s\!+\!m_d)^2+2y^2 v^2} }  \right) ,
\label{eq:12}
\end{equation}
where $y^2=y_1^2+y_2^2$. The first two terms in \eqref{eq:12} are the exact solution of the second order remnant of the characteristic equation  for $m_{\chi}=m_d$ in the  $c_{Z \chi \chi}=0$ blind spot.
The correction is proportional to $x_-$.\footnote{The real expansion parameter for small $c_{Z\chi\chi}$ in equation \eqref{eq:char} is $\frac{m_d \left (y_1 \mp y_2 \right )^2 v^2} { (2m_s^2+6m_d^2+3 y^2 v^2 )^{\frac{3}{2}}}$, which has  positive definite denominator.}
The spectrum in the blind spot $y_1=-y_2$ is shown in Fig. \ref{fig:case1spec}. 

For $m_s < m_d$, the dark matter candidate is the one with mass $m_2$, until $y<\frac{2}{v}\sqrt{m_d(m_d-m_s)}$, then $m_2$ becomes larger than $m_1$. Its couplings are
\begin{equation}
c_{h \chi \chi} =  \frac{8 y^2 v   }{m_s+ m_d},\; \; 
c_{Z \chi \chi} = \frac{ (y^2_1-y_2^2) m_Z v}{2 \left ( m_s^2-m_d^2 \right ) }.
\end{equation}
We see that the $Z$ coupling scales according to the blind spot condition ($y_1+y_2$), but small Higgs coupling requires small overall Yukawa, $y$ and relatively large $m_s+m_d$ compared to the Higgs VEV. 

For $m_d < m_s$ or $y>\frac{2}{v}\sqrt{m_d(m_d-m_s)}$, the dark matter candidate is mostly doublet like and its mass is $m_1$. The couplings are 
\begin{equation}
c_{h \chi \chi} =  \frac{(y_1+y_2)^2 v m_d}{2 m_d(m_d-m_s)- {y^2v^2} } , \;  \; 
c_{Z \chi \chi} = \frac{ (y^2_1-y_2^2) m_Z v}{4 m_d  (m_s-m_d)+2 y^2 v^2}.
\end{equation}
Both couplings go to zero with $(y_1+y_2)^2$ and $(y_1+y_2)$  multiplied with small mass ratios  for non-degenerate mass parameters $|m_s-m_d|>v$.  
This is the most favored region to avoid direct detection even in the presence of sizable Yukawa interactions if they are tuned 
($y_1 + y_2 \simeq 0$).

There is a smooth limit for $m_s=m_d$
\begin{equation}
c_{h \chi \chi} =-\frac{(y_1+y_2)^2 m_d}{ {y^2v} } , \;  \; 
c_{Z \chi \chi} =\frac{ (y^2_1-y_2^2) m_Z}{2 y^2 v}.
\label{eq:case2d}
\end{equation}
The Higgs coupling goes to zero with $\left(y_1 +y_2 \right )^2 $, while the Z coupling scales with ${y_1+y_2}$ as before.

\subsection*{Small $c_{Z \chi \chi}$, $y_1\simeq y_2,$ $\tan{\theta} \simeq 1$}
\begin{figure}[h]
\begin{centering}
\includegraphics[height=3cm]{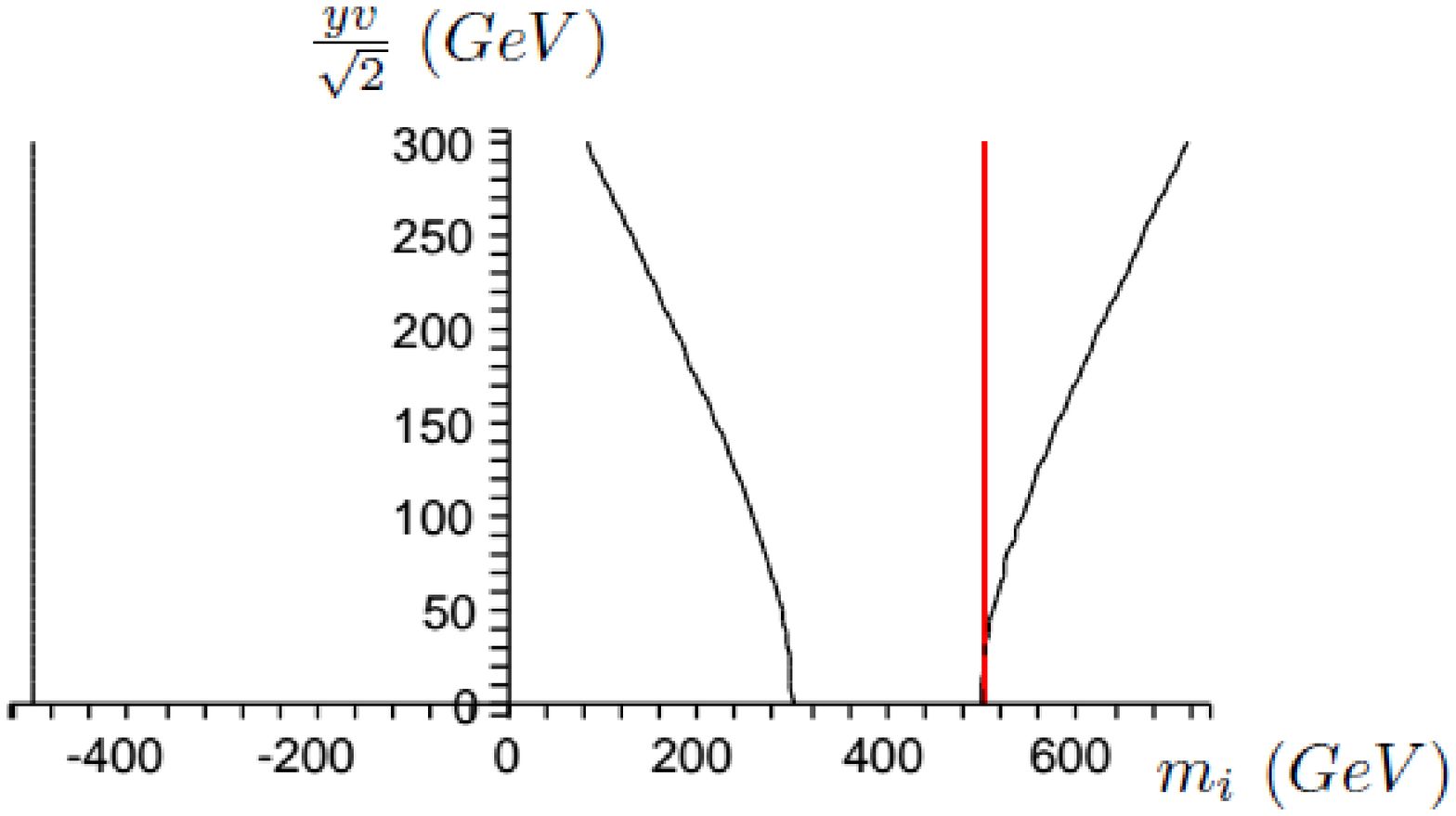}
\includegraphics[height=3cm]{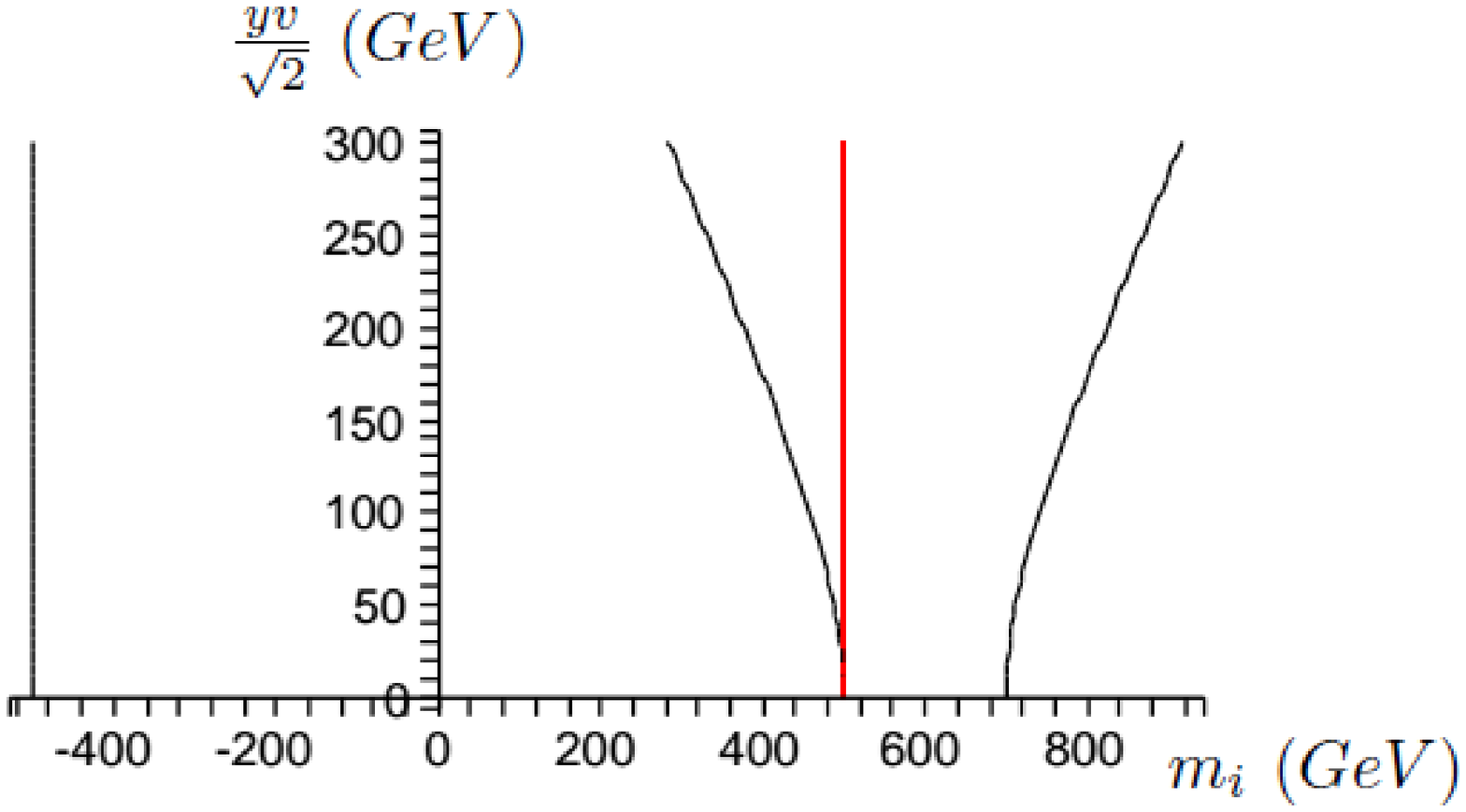}
\par\end{centering}
\centering{}\caption{Mass eigenvalues for $y_1=y_2$ vs. increasing overall Yukawa terms $\frac{yv}{\sqrt{2}}$ for $m_s=300$ GeV  $<m_d=500$ GeV on the left  and $m_s=800$ GeV  $>m_d=500$ GeV on the right. The negative eigenvalue ($m_1$) has been flipped to positive value and both are shown.}
\label{fig:case2spec}
\end{figure}

The mass eigenvalues are
\begin{equation}
m_1=-m_d (1+x_+), \, \quad \mathrm{where} \quad \, x_+ =\frac{ (y_1-y_2)^2 v^2 }{ 4 m_d(m_s+m_d)+y^2{v^2} },
\label{1m1}
\end{equation}

\begin{equation}
m_{2,3}=\frac{m_s+m_d}{2} \pm \frac{1}{2}  \sqrt{(m_s\!-\!m_d)^2+2y^2 v^2}  +
\frac {m_d}{2} x_+ \left(  1 \mp \frac{m_s+3m_d}{\sqrt{ (m_s\!-\!m_d)^2+2y^2 v^2}} \right).
\end{equation}
The spectrum is similar to the first case and it is shown in Fig. \ref{fig:case2spec}.

The lightest fermion with mass $m_3$ is the dark matter candidate. Its mass starts from $m_s$ or $m_d$ and decreases as the Yukawa couplings are increased.

The $Z$ coupling goes to zero with $y_1-y_2$ in agreement with the blind spot condition, assuming $y^2v^2 \ll |m_s^2-m_d^2|$ for the sake of simplicity 
\begin{equation}
c_{Z \chi \chi} = (y^2_1-y_2^2)\frac{m_Z v}{2  (m_d^2-m_s^2)} \; \mathrm{for} \; m_s<m_d ,
\end{equation}
and
\begin{equation}
c_{Z \chi \chi} = (y^2_1-y_2^2)\frac{m_Z v}{2 m_d (m_s-m_d)} \; \mathrm{for} \; m_s>m_d .
\end{equation}
The leading behavior of the Higgs coupling constant for $m_\chi=m_3$ is for $m_s \neq m_d$
\begin{equation}
c_{h \chi \chi} =\frac{y^2v}{|m_d-m_s|}.
\end{equation}
The Higgs coupling  is small only if  the  Yukawa mass correction $ \left( \frac{yv}{\sqrt{2}} \right )$ is smaller than the doublet-singlet mass splitting.

For $m_s=m_d$ the couplings are
\begin{equation}
c_{h \chi \chi} = -y_1 ,\; \mathrm{and} \,\; c_{Z \chi \chi} = \frac{\left ( y_1^2-y_2^2 \right ) m_Z}{ \left ( 2m_d-y_1 v \right )}.
\end{equation}
The $Z$ coupling can be small for tuned Yukawas and if the new vectors are heavier than the $Z$, but the small Higgs coupling needs small $y_1$.

\subsection*{Small ${ c_{h \chi \chi} }$, $ \sin{2 \theta} \simeq -\frac{ m_{\chi}}{ m_d} $ }
\begin{figure}[h]
\begin{centering}
\includegraphics[height=3cm]{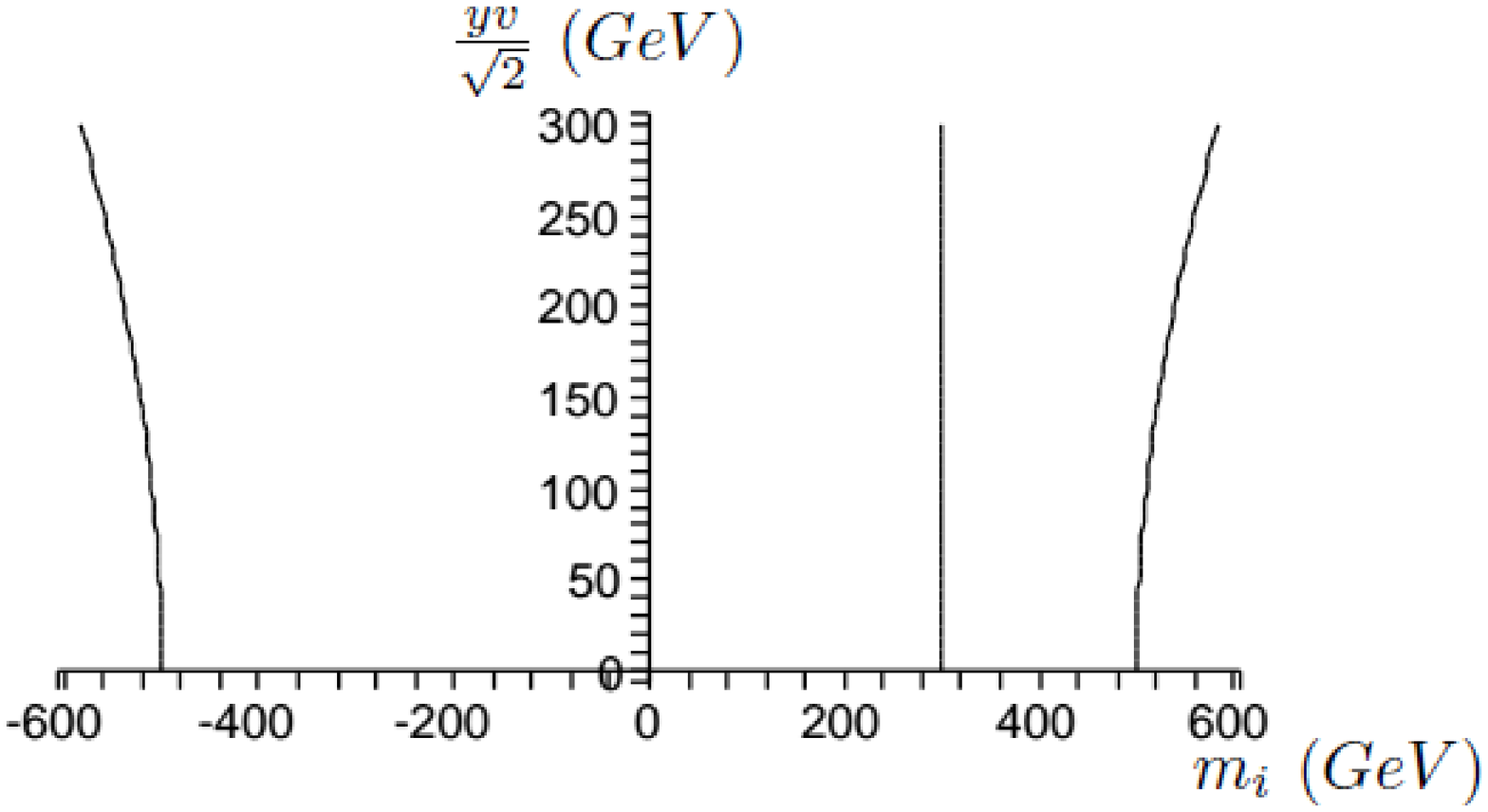}
\includegraphics[height=3cm]{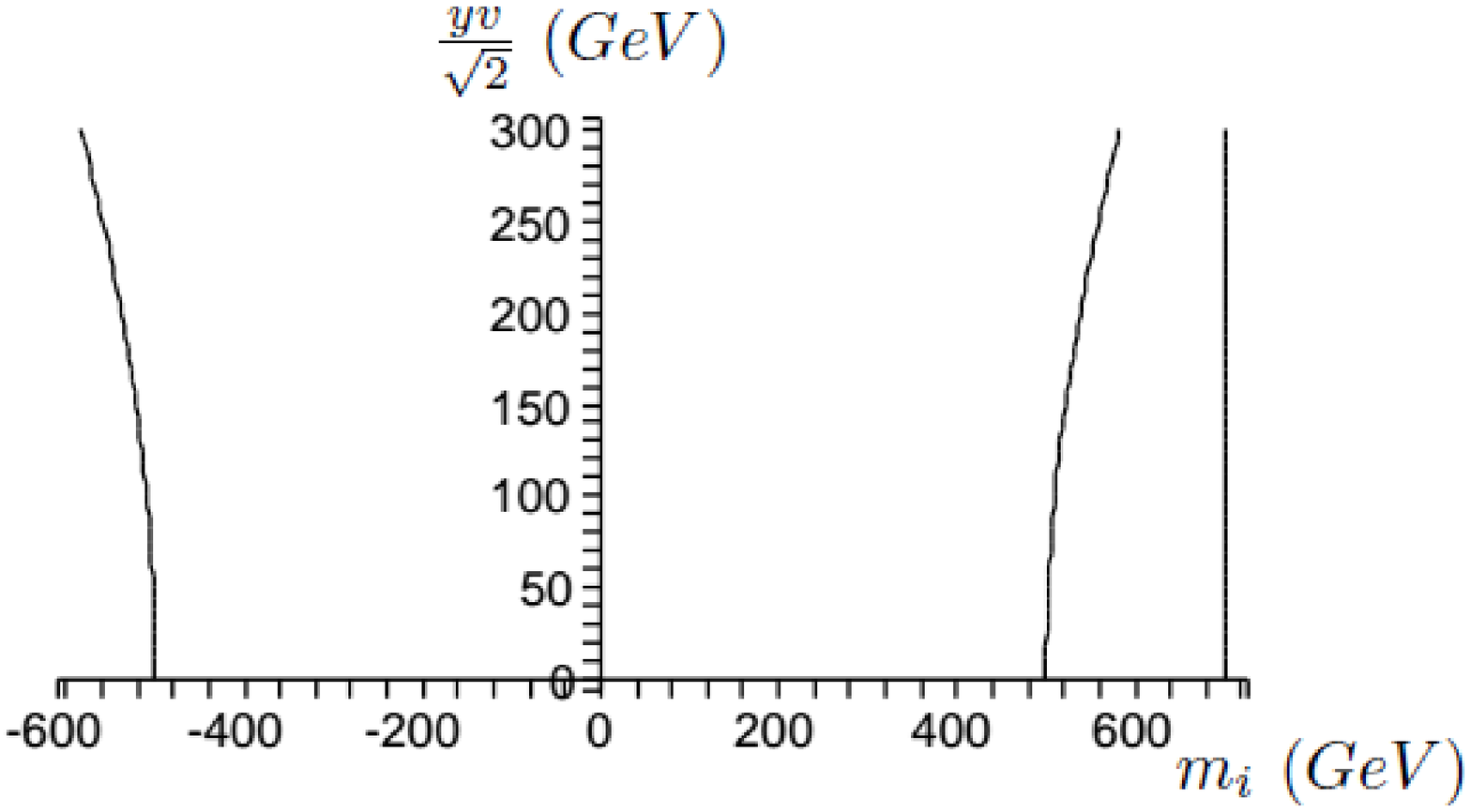}
\par\end{centering}
\centering{}\caption{Mass eigenvalues for $c_{h\chi \chi}=0$ vs. increasing overall Yukawa terms $\frac{yv}{\sqrt{2}}$ for $m_s=300$ GeV  $<m_d=500$ GeV on the left  and $m_s=800$ GeV  $>m_d=500$ GeV on the right. The negative eigenvalue $m_3$ has been flipped to positive value and it coincides with $m_2$.}
\label{fig:case3spec}
\end{figure}

As we mentioned the Higgs blind spot is at $m_\chi=\pm m_d$ or $m_\chi=m_s$.
If $|m_\chi| \simeq m_d$ then $ |\tan \theta|\simeq 1$ and it is discussed in the previous two points. 
The non-trivial new case is when $m_\chi \simeq m_s<m_d$, The mass eigenvalues are
\begin{equation}
m_1=m_s (1-z) ,  \, \quad \mathrm{where} \quad \, z=\frac{ 2 \frac{m_d}{m_s}y_1 y_2 v^2+y^2 v^2 }{2m_d^2-2m_s^2+y^2v^2},
\end{equation}
\begin{equation}
m_{2,3}=\pm \sqrt{m_d^2+\frac{y^2 v^2}{2}}-z\left( \frac{m^2_s}{\sqrt{m_d^2+\frac{y^2 v^2}{2}}} \mp m_s  \right ).
%+\frac {m_d (y_1-y_2)^2 v^2 \left(  1 \mp \frac{m_s+3m_d}{(m_s-m_d)^2+2y^2 v^2} \right) }{2*( 4 m_d(m_s+m_d)-y^2{v^2} )}
\end{equation}
The square root is the exact solution of the second order remnant of the characteristic equation 
in the $c_{h \chi \chi}=0$ blind spot for $m_\chi=m_s$ and the correction is proportional to $z$.\footnote{The expansion parameter for small $c_{h\chi\chi}$ in \eqref{eq:char} is $\frac{( m_s \frac{y^2}{2}+m_d y_1 y_2  )v^2} { (2m_s^2+6m_d^2+3y^2 v^2) ^{\frac{3}{2}}}$.} The spectrum is shown in Fig. \ref{fig:case3spec}.

The coupling constants for the $m_\chi=m_1$ dark matter candidate are
\begin{equation}
c_{h \chi \chi} = 4z \frac{m_s}{v}, \; \;  
c_{Z \chi \chi} = \frac{(y_2^2-y_1^2) m_Z v}{4 m_d \left( m_d- m_s \right )+2y^2v^2}.
\end{equation}
Small $Z$ boson-dark matter coupling requires additional tuning.
The $\frac{y_2}{y_1}$ ratio is fixed by $\frac{m_s}{m_d}$, therefore for non-degenerate masses $m_d-m_s \geq y v$,  $c_{Z \chi \chi}$ scales with $y^2\frac{vm_Z}{m_d \sqrt{m_s^2-m_d^2}}$.
In case of $m_s=m_d$ we get back the first case  and both coupling constants \eqref{eq:case2d} are small.
%$\frac{m_Z v y^2}{m_d \sqrt{m_d^2-m_s^2}}$ must be small, relative light singlet-like DM ($m_s$)  goes with heavier, few TeV $m_d$ mass parameter.

\par We have seen the exclusions and limits in the parameter space, but there are no bounds on the individual parameters. In the following section, we aim to set bounds on the four parameters one by one by exploring the consequences of perturbative unitarity.
%\par There are exclusions in the parameter space from experiments based on numerical studies and analytic expansion. In the following section we explore the consequences of perturbative unitarity and aim to set bounds on single parameters.

\section{Perturbative unitarity} \label{sec:unitarity} % UNITARITY

Perturbative unitarity is an essential tool in exploring effective field theories. 
%The Standard Model gauge interactions can be built up studying the two particle elastic scattering amplitudes. 
Amplitudes growing with energy indicate the breakdown of the effective theory and the validity range of the model can be estimated or new particles and interactions can be added to  cancel the terms with bad high energy behavior. In the SM, the complete gauge structure of the $W$, $Z$ and Higgs interactions can be recovered and there are cancellations due to gauge symmetry. Unitarity still constrains the parameters of the SM, the Higgs self coupling, which can be translated to the Higgs mass \cite{BenLee}.
%The parameters of the model are still constrained after canceling the growing amplitude, the old theoretical upper bound on the Higgs mass is the result of  transferring the upper bound on the Higgs self coupling to the Higgs mass \cite{BenLee}. 
Later, the method was applied to massive chiral fermions without a Higgs boson. It was shown in \cite{appel} that  the scattering amplitude of a fermion-antifermion pair to longitudinally polarized gauge bosons must be unitarized below 3.5 TeV in the case of the top quark, constraining the scale of fermion mass generation, e.g. giving an upper bound on the validity of the model.

Here, we investigate tree-level elastic two-particle scattering processes. Considering the $J=0$ partial-wave amplitude and we require perturbative unitarity for a process with scattering amplitude $\mathcal{M}$ and scattering angle $\theta$.
\begin{equation}
a_0=\frac{1}{32\pi}\intop_{-1}^{1}d\left(\cos{\theta}\right)|\mathcal{M}|\, \, \text{, } \quad |\text{Re}a_0|\leq\frac{1}{2}
\label{unit}
\end{equation}

\par There are four relevant processes in the model. Charged fermion pair annihilation to W's, charged and neutral fermion annihilation to W and Higgs, two charged fermions scattering through Z and $\gamma$ and finally, neutral fermions scattering through Higgs exchange. The amplitudes are included for each helicity channel of the fermions for energies much higher than any masses, $\sqrt{s}\gg m_W$. 

\subsection*{$\Psi^-(s_1)\Psi^+(s_2)\rightarrow W^-W^+$} % FF->WW
\begin{figure}[h]
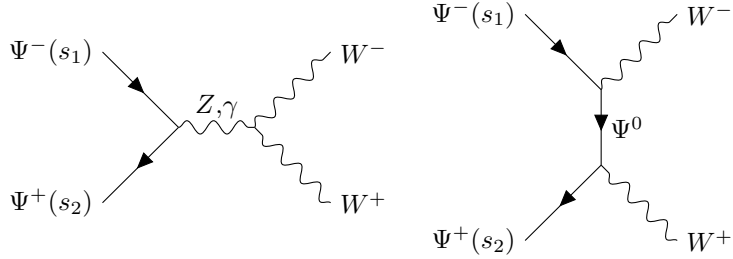

\centering
$\begin{matrix}\ffvvvS{\Psi^-(s_1)}{\Psi^+(s_2)}{W^-}{W^+}{Z{,}\gamma} & 		\fffvvT{\Psi^-(s_1)}{\Psi^+(s_2)}{W^-}{W^+}{\Psi^0}\end{matrix}$
\caption{Feynman graphs for $\Psi^-\Psi^+\rightarrow W^-W^+$ scattering.}
\label{fig:WWscat}
\end{figure}

The Feynman graphs for the process are shown in Fig. \ref{fig:WWscat}. In the t-channel we have three graphs with the three neutral mass eigenstates, $\chi_{1,2,3}$. Since we sum up all possible internal particles, if we do a unitary transformation on $\chi_{1,2,3}$ and sum up all the new states, we get the same result. That means, we can calculate with the electroweak eigenstates, $\Psi^0=\begin{pmatrix}\psi_1^0 \\ \psi_2^{0\dagger}\end{pmatrix}$ and $\chi^0$, where $\chi^0$'s couplings are zero. The amplitude is then
\begin{equation}
\mathcal{M}_{s_1s_2}\left(\Psi^-\Psi^+\rightarrow W^-W^+\right)=
\begin{pmatrix}
\frac{g^2(1-\tan^2{\theta_w})}{4}\sin{\theta}+\mathcal{O}\left(\frac{m_W^2}{s}\right) & 
\mathcal{O}\left(\frac{m_W}{\sqrt{s}}\right) \\ 
\mathcal{O}\left(\frac{m_W}{\sqrt{s}}\right) & 
-\frac{g^2(1-\tan^2{\theta_w})}{4}\sin{\theta}+\mathcal{O}\left(\frac{m_W^2}{s}\right)
\end{pmatrix}
\end{equation}
In the amplitude matrix, the incoming fermion helicity channels are the following, 
\begin{equation} \label{eq:2hel}
\begin{pmatrix}s_1 & s_2\end{pmatrix}:\, \begin{pmatrix} -- & -+ \\ +- & ++\end{pmatrix}.
\end{equation}
\par As in the case of chiral fermions \cite{appel}, the s-channel amplitudes grow with $s$, that cancel with the $Z$ and $\gamma$ exchange graphs. The t-channel grows with $\sqrt{s}$, but here it is canceled by the s-channel graphs and there is no need for the otherwise absent Higgs exchange.

%\subsection*{$\Psi^-(s_1)\chi^0(s_2)\rightarrow W^-h$ and $\Psi^-(s_1)\Psi^0(s_2)\rightarrow W^-h$} 
\subsection*{$\Psi^-(s_1)\chi_i(s_2)\rightarrow W^-h$} % FF->WH
\begin{figure}[h]
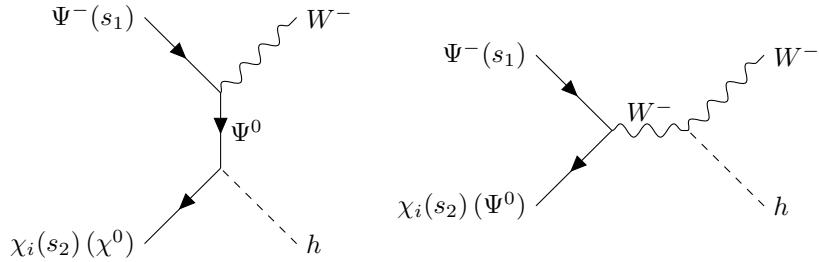

\centering
$\begin{matrix}\fffvhT{\Psi^-(s_1)}{\chi_i(s_2)\,(\chi^0)}{W^-}{\Psi^0} & \ffvvhS{\Psi^-(s_1)}{\chi_i(s_2)\,(\Psi^0)	}{W^-}{W^-}\end{matrix}$
\caption{Feynman graphs for the $\Psi^-\chi_i\rightarrow W^-h$ scattering.}
\label{fig:WHscat}
\end{figure}

The charged and a neutral fermion can annihilate into a $W$ and Higgs, illustrated in Fig.  \ref{fig:WHscat}. 
In the t-channel internal lines, we can use again the weak eigenstate $\Psi^0$. Naively using the weak eigenstates as the initial states, we find that the $\Psi^-\chi^0\rightarrow W^-h$ scattering process grows with $\sqrt{s}$, while there is no problem with the other process $\Psi^-\Psi^0\rightarrow W^-h$ (the helicities of the fermions are as in \eqref{eq:2hel}).
\begin{equation}
\mathcal{M}_{s_1s_2}\left(\Psi^-\chi^0\rightarrow W^-h\right)=\begin{pmatrix}
0 & \frac{gy_2\sqrt{s}}{2m_W}+\mathcal{O}\left(\frac{m_W}{\sqrt{s}}\right) \\
-\frac{gy_1\sqrt{s}}{2m_W}+\mathcal{O}\left(\frac{m_W}{\sqrt{s}}\right) & 0
\end{pmatrix}
\end{equation}
\begin{equation}
\mathcal{M}_{s_1s_2}\left(\Psi^-\Psi^0\rightarrow W^-h\right)=\begin{pmatrix}
-\frac{g^2\sin{\theta_w}}{2\sqrt{2}} & 0 \\
0 & \frac{g^2\sin{\theta_w}}{2\sqrt{2}}
\end{pmatrix}+\mathcal{O}\left(\frac{m_W}{\sqrt{s}}\right)
\end{equation}
But taking the mass eigenstates $\chi_{1,2,3}$, that mixes $\Psi^0$ and $\chi^0$, makes the s-channel graph grows with $\sqrt{s}$ compensating the t-channel. The whole process becomes unitary and leaving the highest order to be proportional to $g^2\sin{\theta_w}$ and a combination of the mixing matrix elements.

\subsection*{$\Psi^-(s_1)\Psi^+(s_2)\rightarrow\Psi^-(s_3)\Psi^+(s_4)$} % FF-Z/G-FF
\begin{figure}[h]
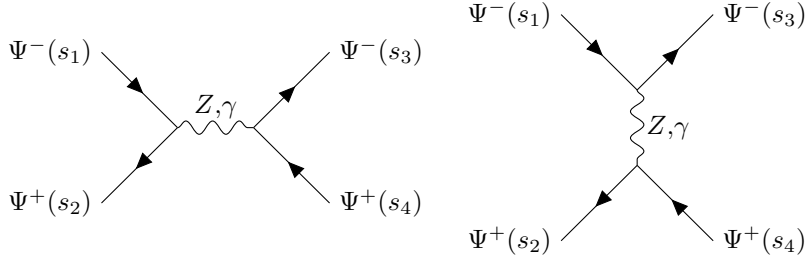

\centering
$\begin{matrix}\ffvffS{\Psi^-(s_1)}{\Psi^+(s_2)}{\Psi^-(s_3)}{\Psi^+(s_4)}{Z{,}\gamma} & \ffvffT{\Psi^-(s_1)}{\Psi^+(s_2)}{\Psi^-(s_3)}{\Psi^+(s_4)}{Z{,}\gamma}\end{matrix}$
\caption{Feynman graphs for $\Psi^-\Psi^+\rightarrow\Psi^-\Psi^+$ scattering.}
\label{fig:PPcharged}
\end{figure}

The charged fermions can scatter through s- and t-channel $Z$ and $\gamma$ exchange illustrated with Feynman graphs in Fig. \ref{fig:PPcharged}.
\begin{equation}
\mathcal{M}_{s_1s_2s_3s_4}\left(\Psi^-\Psi^+\rightarrow\Psi^-\Psi^+\right)=\begin{pmatrix}
\frac{g^2\sin^2{\frac{\theta}{2}}}{2\cos^2{\theta_w}} & 0 & 0 &
\frac{g^2\cos^2{\frac{\theta}{2}}}{2\cos^2{\theta_w}} \\ 
0 & 0 & \frac{-g^2}{2\cos^2{\theta_w}} & 0 \\
0 & \frac{-g^2}{2\cos^2{\theta_w}} & 0 & 0 \\
\frac{g^2\cos^2{\frac{\theta}{2}}}{2\cos^2{\theta_w}} & 0 & 0 &
\frac{g^2\sin^2{\frac{\theta}{2}}}{2\cos^2{\theta_w}}
\end{pmatrix}+\mathcal{O}\left(\frac{m_W^2}{s}\right)
\end{equation}
For the two-to-two fermion scattering amplitudes the sixteen fermion helicity channels are in the following order, 
\begin{equation} \label{eq:helicity}
\begin{pmatrix} s_1 & s_2 & s_3 & s_4 \end{pmatrix}:\, 
\begin{pmatrix}
---- & ---+ & -+-- & -+-+ \\ --+- & --++ & -++- & -+++ \\ 
+--- & +--+ & ++-- & ++-+ \\ +-+- & +-++ & +++- & ++++ \\ 
\end{pmatrix}.
\end{equation}

In the first two processes, the growing terms in the amplitude cancel each other because of  the gauge symmetry. The remaining term, just as in the previous two processes, is proportional to the gauge coupling and satisfy the unitarity constraint \eqref{unit}.

\subsection*{$\Psi^0(s_1)\chi^0(s_2)\rightarrow\Psi^0(s_3)\chi^0(s_4)$} % FF-H-FF
\begin{figure}[h]
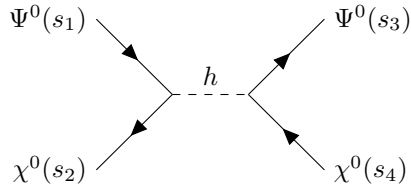

\centering
$\ffhffS{\Psi^0(s_1)}{\chi^0(s_2)}{\Psi^0(s_3)}{\chi^0(s_4)}$
\caption{Feynman graphs for $\Psi^0\chi^0\rightarrow\Psi^0\chi^0$ scattering.}
\label{fig:PPneutr}
\end{figure}

The neutral fermion scattering can be still interesting, since this process  includes Higgs boson exchange and the corresponding unconstrained Yukawa couplings. The related Feynman graph is shown in Fig. \ref{fig:PPneutr}. This process is basically the $\chi_i\chi_j\rightarrow\chi_k\chi_l$ mass eigenstate scattering. But since there is only one type of graph contributing, which is proportional to some combination of the Yukawa couplings, $y_{1,2}$, we can mix these processes to get the $\Psi^0\chi^0\rightarrow\Psi^0\chi^0$ weak eigenstate scattering with the following amplitude.
\begin{multline}
i\mathcal{M}\left(\Psi^0_{s_1}(p_1)\chi^0_{s_2}(p_2)\rightarrow h(q_s)\rightarrow\Psi^0_{s_3}(p_3)\chi^0_{s_4}(p_4)\right)= \\
=\left(\frac{i}{2\sqrt{2}}\right)^2 \left(\bar{v}_{s_2}(p_2)\left(y_++y_-\gamma_5\right)u_{s_1}(p_1)\right)\frac{i}{s-m_h^2}\left(\bar{u}_{s_3}(p_3)\left(y_++y_-\gamma_5\right)v_{s_4}(p_4)\right)
\end{multline}
Where the shorthand notations of the  couplings are $y_{\pm}=y_1\pm y_2$. 
\begin{equation}
\mathcal{M}_{s_1s_2s_3s_4}\left(\Psi^0\chi^0\rightarrow\Psi^0\chi^0\right)=\begin{pmatrix}
0 & 0 & 0 & 0 \\
0 & \frac{y_2^2}{2}+\mathcal{O}\left(\frac{m_h^2}{s}\right) & -\frac{y_1y_2}{2}+\mathcal{O}\left(\frac{m_h^2}{s}\right) & 0 \\
0 & -\frac{y_1y_2}{2}+\mathcal{O}\left(\frac{m_h^2}{s}\right) & \frac{y_1^2}{2}+\mathcal{O}\left(\frac{m_h^2}{s}\right) & 0 \\
0 & 0 & 0 & 0 \\
\end{pmatrix}
\end{equation}
The fermion helicity channels are in the same order as in the charged fermion scattering in \eqref{eq:helicity}.
\par Now using $a_0$ partial-wave unitarity \eqref{unit} for the non-zero elements in the amplitude matrix, we get a bound on the Yukawa couplings. 
\begin{equation}
\begin{matrix}
|y_{1,2}|\leq 4\sqrt{\pi}\approx 7.1 & |y_1y_2|\leq 16\pi\approx 50.3
\end{matrix}
\end{equation}
The Yukawa contributions to the fermion masses are bounded by
\begin{equation}
\frac{|y_{1,2}|v}{\sqrt{2}}\leq 2\sqrt{2\pi}v\approx 1.23 \, \text{TeV,}
\end{equation}
where $v=246 \, \text{TeV}$ is the Higgs VEV.
\par We see that perturbative unitarity  sets meaningful bounds on the so far unconstrained new Yukawa couplings. The other processes are proportional to the  gauge coupling and  the bounds are automatically satisfied.

\section{Conclusion} \label{sec:conlcusion} % CONCLUSION
We have studied the vector-like fermionic singlet-doublet extension of the Standard Model. This is a minimal, gauge invariant and renormalizable model, motivated by dark matter. 
It has four free parameters, two masses and two Yukawa couplings.
A $\mathbb{Z}_2$ matter parity is introduced, the lightest neutral mass eigenstate can be the dark matter candidate.
As a result of the mixing between the doublet and the singlet there are tree-level couplings of the dark matter to the $Z$ and Higgs bosons and the couplings can vary in a wide range. This model is a consistent simplified version of UV complete supersymmetric theories, such as the bino-higgsino sector of the MSSM or the singlino-higgsino sector of the naturalness motivated NMSSM-like models. The main difference from the MSSM bino-higgsino system is that the Yukawa couplings are free parameters here and not related to the the hypercharge coupling and $\tan \beta$.  The effective or simplified model is ideal to test the parameters with experiments. 

To have a better view and analytic control of the parameters, we have calculated the mass eigenvalues and the relevant $Z, \, h$ couplings in the neighborhood of the blind spots. 
We  identified the best region allowed by direct detection experiments, where $ m_d < m_s $ and the not necessary small Yukawas are tuned to nearly cancel each other $y_1+y_2 \simeq 0$. There are other blind spot regions, where the Yukawas ($y$) must be small. 
Direct and indirect dark matter searches constrain the combined parameter space, but not the individual parameters of the model, still leaving room for a "WIMP miracle".  To set bounds on separate parameters, we have calculated two particle scattering amplitudes in the model.
Applied the bounds of  perturbative unitarity and found that no amplitude has bad high energy behavior as expected from renormalizability. There are no constraints on the  Dirac- and Majorana-mass. The new Yukawa couplings appear in s-channel Higgs exchange graphs and are bounded $|y_{1,2}|<7.1$. The LHC phenomenology of the model is studied e.g. in \cite{survey,SDbounds}. The medium and the low mass region can be tested at the $\sqrt{s}= 13$ TeV LHC in the next few years \cite{SDbounds}.

%\newpage


\begin{thebibliography}{9}
% 1
\bibitem{survey}
  S.~A.~R.~Ellis, R.~M.~Godbole, S.~Gopalakrishna and J.~D.~Wells,
  %``Survey of vector-like fermion extensions of the Standard Model and their phenomenological implications,''
  JHEP {\bf 1409} (2014) 130
  [arXiv:1404.4398 [hep-ph]].
  
\bibitem{cynprec}
  G.~Cynolter and E.~Lendvai,
  %``Electroweak Precision Constraints on Vector-like Fermions,''
  Eur.\ Phys.\ J.\ C {\bf 58} (2008) 463
  [arXiv:0804.4080 [hep-ph]].

\bibitem{schwaller}
  A.~Joglekar, P.~Schwaller and C.~E.~M.~Wagner,
  %``Dark Matter and Enhanced Higgs to Di-photon Rate from Vector-like Leptons,''
  JHEP {\bf 1212} (2012) 064
  [arXiv:1207.4235 [hep-ph]].
  
\bibitem{UED}
  T.~Appelquist, H.~C.~Cheng and B.~A.~Dobrescu,
  %``Bounds on universal extra dimensions,''
  Phys.\ Rev.\ D {\bf 64} (2001) 035002
  [hep-ph/0012100]. 

\bibitem{lH}
  N.~Arkani-Hamed, A.~G.~Cohen, E.~Katz and A.~E.~Nelson,
  %``The Littlest Higgs,''
  JHEP {\bf 0207} (2002) 034
  [hep-ph/0206021].
% 5
\bibitem{lH2}
  N.~Arkani-Hamed, A.~G.~Cohen, E.~Katz, A.~E.~Nelson, T.~Gregoire and J.~G.~Wacker,
  %``The Minimal moose for a little Higgs,''
  JHEP {\bf 0208} (2002) 021
  [hep-ph/0206020].

\bibitem{lowSUSY}
  N.~Arkani-Hamed and S.~Dimopoulos,
  %``Supersymmetric unification without low energy supersymmetry and signatures for fine-tuning at the LHC,''
  JHEP {\bf 0506} (2005) 073
  [hep-th/0405159].

\bibitem{Contino}
  R.~Contino, L.~Da Rold and A.~Pomarol,
  %``Light custodians in natural composite Higgs models,''
  Phys.\ Rev.\ D {\bf 75} (2007) 055014 
  [hep-ph/0612048].

\bibitem{fcm} 
G.~Cynolter, E.~Lendvai and G.~Pocsik,
  %``Fermion condensate model of electroweak interactions,''
  Eur.\ Phys.\ J.\ C {\bf 46} (2006) 545
  [hep-ph/0509230].

\bibitem{DMandUnif}
  R.~Mahbubani and L.~Senatore,
  %``The Minimal model for dark matter and unification,''
  Phys.\ Rev.\ D {\bf 73} (2006) 043510
  [hep-ph/0510064].
% 10
\bibitem{Hall} 
  R.~Enberg, P.~J.~Fox, L.~J.~Hall, A.~Y.~Papaioannou and M.~Papucci,
  %``LHC and dark matter signals of improved naturalness,''
  JHEP {\bf 0711} (2007) 014
  [arXiv:0706.0918 [hep-ph]].

\bibitem{DEramo}
  F.~D'Eramo,
  %``Dark matter and Higgs boson physics,''
  Phys.\ Rev.\ D {\bf 76} (2007) 083522
  [arXiv:0705.4493 [hep-ph]].

\bibitem{ligeti}
  K.~Ishiwata, Z.~Ligeti and M.~B.~Wise,
  %``New Vector-Like Fermions and Flavor Physics,''
  arXiv:1506.03484 [hep-ph].

\bibitem{SDbounds}
  L.~Calibbi, A.~Mariotti and P.~Tziveloglou,
  %``Singlet-Doublet Model: Dark matter searches and LHC constraints,''
  arXiv:1505.03867 [hep-ph].
% 15 
\bibitem{BenLee}
  B.~W.~Lee, C.~Quigg and H.~B.~Thacker,
  %``Weak Interactions at Very High-Energies: The Role of the Higgs Boson Mass,''
  Phys.\ Rev.\ D {\bf 16} (1977) 1519.

\bibitem{appel}
  T.~Appelquist and M.~S.~Chanowitz,
  %``Unitarity Bound on the Scale of Fermion Mass Generation,''
  Phys.\ Rev.\ Lett.\  {\bf 59} (1987) 2405
  [Phys.\ Rev.\ Lett.\  {\bf 60} (1988) 1589].

\bibitem{singletdoublet}
  T.~Cohen, J.~Kearney, A.~Pierce and D.~Tucker-Smith,
  %``Singlet-Doublet Dark Matter,''
  Phys.\ Rev.\ D {\bf 85} (2012) 075003
  [arXiv:1109.2604 [hep-ph]].

\bibitem{Higgsoasis}
  C.~Cheung, M.~Papucci and K.~M.~Zurek,
  %``Higgs and Dark Matter Hints of an Oasis in the Desert,''
  JHEP {\bf 1207} (2012) 105
  [arXiv:1203.5106 [hep-ph]].

\bibitem{simplModels}
  C.~Cheung and D.~Sanford,
  %``Simplified Models of Mixed Dark Matter,''
  JCAP {\bf 1402} (2014) 011
  [arXiv:1311.5896 [hep-ph]].
% 20  
\bibitem{Iceberg}
  J.~Halverson, N.~Orlofsky and A.~Pierce,
  %``Vectorlike Leptons as the Tip of the Dark Matter Iceberg,''
  Phys.\ Rev.\ D {\bf 90} (2014) 1,  015002
  [arXiv:1403.1592 [hep-ph]].

\bibitem{Planck}
  P.~A.~R.~Ade {\it et al.}  [Planck Collaboration],
  %``Planck 2013 results. XVI. Cosmological parameters,''
  Astron.\ Astrophys.\  {\bf 571} (2014) A16
  [arXiv:1303.5076 [astro-ph.CO]].
  
\bibitem{FermiLAT}
  M.~Ackermann {\it et al.}  [Fermi-LAT Collaboration],
  %``Searching for Dark Matter Annihilation from Milky Way Dwarf Spheroidal Galaxies with Six Years of Fermi-LAT Data,''
  arXiv:1503.02641 [astro-ph.HE].

\bibitem{IceCube} 
  M.~G.~Aartsen {\it et al.} [IceCube Collaboration],
  %``Search for dark matter annihilations in the Sun with the 79-string IceCube detector,''
  Phys.\ Rev.\ Lett.\  {\bf 110} (2013) 13,  131302
  [arXiv:1212.4097 [astro-ph.HE]]. 

\bibitem{PICO}
  C.~Amole {\it et al.}  [PICO Collaboration],
  %``Dark Matter Search Results from the PICO-2L C$_3$F$_8$ Bubble Chamber,''
  Phys.\ Rev.\ Lett.\  {\bf 114} (2015) 23,  231302
  [arXiv:1503.00008 [astro-ph.CO]].
% 25
\bibitem{LUX}
  D.~S.~Akerib {\it et al.}  [LUX Collaboration],
  %``First results from the LUX dark matter experiment at the Sanford Underground Research Facility,''
  Phys.\ Rev.\ Lett.\  {\bf 112} (2014) 091303
  [arXiv:1310.8214 [astro-ph.CO]].

\bibitem{X100}
  E.~Aprile {\it et al.} [XENON100 Collaboration],
  %``Limits on spin-dependent WIMP-nucleon cross sections from 225 live days of XENON100 data,''
  Phys.\ Rev.\ Lett.\  {\bf 111} (2013) 2,  021301
  [arXiv:1301.6620 [astro-ph.CO]].
  
\bibitem{LEP}
  J.~Abdallah {\it et al.}  [DELPHI Collaboration],
  %``Searches for supersymmetric particles in e+ e- collisions up to 208-GeV and interpretation of the results within the MSSM,''
  Eur.\ Phys.\ J.\ C {\bf 31} (2003) 421
  [hep-ex/0311019].
  
\bibitem{LHC}
  R.~Enberg, P.~J.~Fox, L.~J.~Hall, A.~Y.~Papaioannou and M.~Papucci,
  %``LHC and dark matter signals of improved naturalness,''
  JHEP {\bf 0711} (2007) 014
  [arXiv:0706.0918 [hep-ph]].
  
\bibitem{blindspot}
  C.~Cheung, L.~J.~Hall, D.~Pinner and J.~T.~Ruderman,
  %``Prospects and Blind Spots for Neutralino Dark Matter,''
  JHEP {\bf 1305} (2013) 100
  [arXiv:1211.4873 [hep-ph]].
  
\end{thebibliography}
\end{document}